\documentclass[a4paper,11pt]{article}
\usepackage{pos}

\newcommand{\beq}{\begin{eqnarray}}
\newcommand{\eeq}{\end{eqnarray}}

\usepackage{bbm}

\title{Bump of sound velocity in dense 2-color QCD}

\author*[a,b,c]{Etsuko Itou}
\author[d]{Kei Iida}

\affiliation[a]{\it {Interdisciplinary Theoretical and Mathematical Sciences Program (iTHEMS), RIKEN, Wako 351-0198, Japan}}
\affiliation[b]{\it {Department of Physics, and Research and 
Education Center for Natural Sciences, Keio University, 4-1-1 Hiyoshi, Yokohama, Kanagawa 223-8521, Japan}}
\affiliation[c]{\it {Research Center for Nuclear Physics (RCNP), Osaka University, Osaka 567-0047, Japan}}
\affiliation[d]{{\it
Department of Mathematics and Physics, Kochi University, 2-5-1 Akebono-cho, Kochi 780-8520, Japan
}}

\emailAdd{itou@yukawa.kyoto-u.ac.jp}

\abstract{We obtain the equation of state (EoS) and the sound velocity for 2-color QCD at low temperature and high density and find that in the superfluid phase, $c_s^2/c^2 >1/3$, where $1/3$ is the value at the relativistic limit. Several independent Monte Carlo studies on 2-color QCD have been conducted intensively in recent years. These works have shown a clear evidence of phase transition between hadronic and superfluid phases.
In our paper~\cite{Iida:2022hyy}, we have investigated the EoS and sound velocity in both phases.
Our result is consistent with chiral perturbation theory in a low $\mu$ regime of the superfluid phase including the Bose-Einstein condensed phase, and shows a peak of sound velocity in the high-density BCS phase.
We also give detailed simulation results and a comment on the holography bound in this proceedings.
}

\FullConference{%
The 39th International Symposium on Lattice Field Theory,\\
8th-13th August, 2022,\\
Rheinische Friedrich-Wilhelms-Universit\"{a}t Bonn, Bonn, Germany
}

\begin{document}
\maketitle

\section{Introduction}
Determination of the equation of state (EoS) for dense QCD at low temperature has been desired recently especially because it is related with understanding neutron star observations including recent simultaneous measurements of masses and radii of neutron stars.
However, the first-principles calculation of dense QCD at low temperature, beyond the onset scale ($\mu/m_{\pi} >1/2$) in particular, is still extremely difficult because of the severe sign problem.
On the other hand, the sign problem is absent in even-flavor dense $2$-color QCD  because of the pseudo-reality of fundamental quarks.
Furthremore, if we add  an external source term of the diquark condensate to explicitly break the U(1) baryon symmetry, we can perform numerical simulations using an exact algorithm even beyond the onset scale, namely, in the superfluid phase.
$2$-color QCD at zero chemical potential exhibits the same properties  as $3$-color QCD, {\it e.g.,} confinement, spontaneous chiral symmetry breaking, and 
thermodynamic behaviors.
It is expected that $2$-color QCD even at non-zero chemical potential  could be a good testing ground  in qualitatively understanding dense QCD.

Based on this motivation, several Monte Carlo studies on $2$-color QCD have been conducted  independently and intensively in recent years (see references in Ref.~\cite{Iida:2022hyy}).
One can conclude that the $2$-color QCD phase diagram has been quantitatively clarified; even at fairly high temperature, $T\approx 100$ MeV, superfluidity can remain.

Now, we would like to focus on the EoS and the sound velocity in a low temperature and high density regime.
Several early works based on a phenomenological quark-hadron crossover picture of neutron star matter~\cite{Masuda2013-jk,Baym:2017whm} suggested that
 the  zero-temperature sound velocity squared, $c_s^2=\partial p/\partial e$, peaks in $n_B = 1$--$10n_o$ to be consistent with various observational constraints.
Here, $p$, $e$ and $n_0$ denote the pressure, internal energy density of the system, and nuclear saturation density, respectively.
More recently, based on a quarkyonic matter model, McLerran and Reddy~\cite{McLerran2019-qh} have shown that the peak appears at $n_B= 1$--$5n_0$.
Furthermore, Kojo ~\cite{Kojo2021-mg} proposed a microscopic interpretation on the origin of the peak 
based on a quark saturation mechanism, which is supposed to work for any number of colors. Actually, Kojo and Suenaga~\cite{Kojo2021-wh} argued that a similar peak  of $c_s^2$ emerges not only in $3$-color QCD, but also in $2$-color QCD.

\section{Lattice setup}
The lattice gauge action used in this work is the Iwasaki gauge action
As for the fermion action, we take the naive Wilson fermion with the quark number density and diquark source terms,
\beq
S_F&=& (\bar{\psi_{1}} ~~ \bar{\varphi}) \left( 
\begin{array}{cc}
\Delta(\mu) & J \gamma_5 \\
-J \gamma_5 & \Delta(-\mu) 
\end{array}
\right)
\left( 
\begin{array}{c}
\psi_{1}  \\
\varphi  
\end{array}
\right)
 \equiv  \bar{\Psi} {\mathcal M} \Psi, \nonumber\\ \label{eq:def-M}
\eeq
where
$\bar{\varphi}=-\psi_2^T C \tau_2, ~~~ \varphi=C^{-1} \tau_2 \bar{\psi}_2^T.$
Here, the indices $1,2$ of $\psi$ denote the label of the flavor, and the $\Delta(\mu)_{x,y}$  is the Wilson-Dirac operator with the number operator.
The additional parameter $J$ corresponds to the diquark source parameter,
which allows us to perform the numerical simulation in the superfluid phase.
Note that $J=j \kappa$, where $j$ is a source parameter in the corresponding continuum theory, and $\kappa$ is the hopping parameter.
The $C$ in $\bar{\varphi},\varphi$ is the charge conjugation operator, and $\tau_2$ acts on the color index.
The square of the extended matrix ($\mathcal M$) can be diagonal, 
but $\det[{\mathcal M}^\dag {\mathcal M}]$ corresponds to the fermion action for the four-flavor theory, since a single $\mathcal{M}$ in Eq.\ (\ref{eq:def-M})  represents the fermion kernel of the two-flavor theory.
To reduce the number of fermions, we take the root of the extended matrix in the action.
In practice, utilizing the Rational Hybrid Monte Carlo (RHMC) algorithm, we can generate gauge configurations.

In this work, we perform the simulation with $(\beta, \kappa,N_s,N_\tau)=(0.80,0.159,16,16)$.
According to Ref.~\cite{Iida:2020emi}, once we introduce the physical scale as $T_c=200$ MeV, where $T_c$ denotes the pseudo-critical temperature of chiral phase transition at $\mu=0$, then  our parameter set, $\beta=0.80$ and $N_\tau=16$ ($T=0.39T_c$), corresponds to $a\approx 0.17$ fm and $T\approx 79$ MeV.
The mass of the lightest pseudo-scalar (PS) meson at $\mu=0$, $m_{PS}$, is still heavy in our simulations, $am_{PS}=0.6229(34)$ ($m_{PS}\approx 750 $ MeV).
As for the values of $a\mu$, we generate the configurations at intervals of $a\Delta \mu=0.05$. 
The  number of configuration for each parameter is $100$--$300$. 
The statistical errors are estimated by the jackknife method.

\section{Phase structure at $T=79$ MeV}

\begin{figure}[htbp]
 \begin{center}
    \begin{tabular}{c}
        \includegraphics[keepaspectratio, scale=0.25]{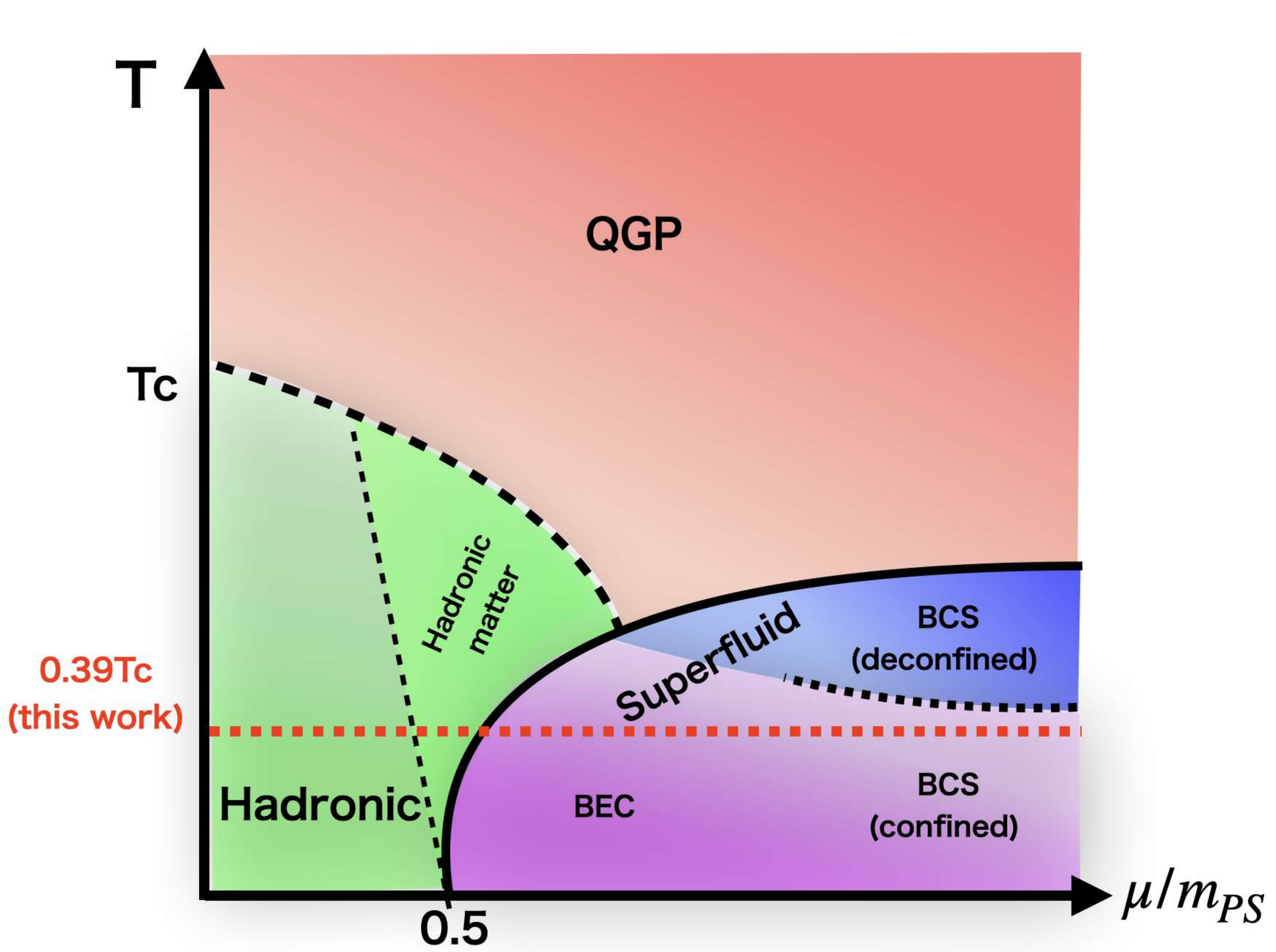}
    \end{tabular}
            \caption{
Schematic 2-color QCD phase diagram.  Each phase is defined in Table \ref{table:phase}.
}\label{fig:phase-diagram}
\end{center}
  \end{figure}

We show the schematic 
phase structure in Fig.~\ref{fig:phase-diagram} and summarize the definition of each phase in Table~\ref{table:phase}, which is  an extract from Ref.~\cite{Iida:2019rah}. 
\begin{table}[h]
\begin{center}
\begin{tabular}{|c||c|c|c|c|}
\hline
 \multicolumn{1}{|c||}{}  & \multicolumn{2}{c|}{Hadronic} &     \multicolumn{2}{c|}{Superfluid}  \\  
\cline{3-3} \cline{4-5}  & & Hadronic matter &  BEC & BCS \\  
 \hline \hline
$\langle |L| \rangle$ & zero  & zero  &    &   \\
$\langle qq \rangle$ & zero  &  zero  & non-zero & non-zero  \\ 
$ \langle n_q \rangle $ &zero &  non-zero & $  0 < \frac{\langle n^{latt.}_q \rangle}{n_q^{\mbox{tree}}} <1 $  & $  \frac{\langle n^{latt.}_q \rangle}{n_q^{\mbox{tree}}} \approx 1 $ \\ 
 \hline
\end{tabular}
\caption{ Definition of phases. } \label{table:phase}
\end{center}
\end{table}
The order parameters  that help classify the  phases are the Polyakov loop $\langle |L| \rangle$ and diquark condensate $\langle qq \rangle$, whose zero/nonzero values indicate the appearance of confinement and superfluidity, respectively.
We found that the superfluidity emerges at $\mu_c/m_{PS} \approx 0.5$ as predicted by the chiral perturbation theory (ChPT)~\cite{Kogut2000-so}.
It is natural to use $\mu/m_{PS}$ as a dimensionless parameter of density since the critical value $\mu_c$ can be approximated by $m_{PS}/2$  even if  the value of $m_{PS}$ in numerical simulation would be changed~\footnote{It is expected that the corresponding critical value of $\mu$ would be $\mu_c/m_N \approx 1/3$ if the hadronic-superfluid phase transition occurs also in the case of $3$-color QCD, where $m_N$ denotes the nucleon mass. }.
We also confirmed that the scaling law of the order parameter around it is consistent with the ChPT prediction.
Furthermore, we measured the quark number operator, $n_q^{latt.}\equiv a^3 n_q= \sum_{i} \kappa \langle \bar{\psi}_i (x) (\gamma_0 -\mathbb{I}_4) e^\mu U_{4} (x) \psi_i (x+\hat{4})  + \bar{\psi}_i (x) (\gamma_0 + \mathbb{I}_4) e^{-\mu}U_4^\dag (x-\hat{4} )\psi_i (x-\hat{4})\rangle$.
We identified the regime where $\langle n^{latt.}_q \rangle$ is consistent with the free quark theory $n_q^{\mathrm{tree}}$ (see Eq.~(26) in Ref.~\cite{Hands2006-mh}) as the BCS phase. 
Thus, we concluded that there are hadronic, hadronic-matter, Bose-Einstein condesed (BEC) and BCS phases at $T=79$ MeV, although there is no clear boundary between the BEC and BCS phases.  Interestingly,
up to $\mu/m_{PS} =1.28 $ ($\mu \lesssim 960$ MeV), the confining behavior remains~\cite{Ishiguro:2021yxr},  while nontrivial instanton configurations have been discovered from calculations of the topological susceptibility~\cite{Iida:2019rah}.
It indicates that  a naive perturbative picture, for instance, pQCD, is not yet valid in the density regime studied here.

\section{Equation of state and velocity of sound at finite $\mu$}
Now, we utilize a fixed scale method to obtain the EoS  at finite density~\cite{Hands2006-mh}.
The trace anomaly can be described by the beta-functions of various parameters and the trace part of the energy-momentum tensor. In our lattice setup, which is explicitly given by
\beq
e-3p &=& \frac{1}{N_s^3 N_\tau} \left( a \frac{d \beta}{da} |_{\mathrm{LCP}} \langle \frac{\partial S}{\partial \beta}\rangle_{sub.}  + a \frac{d \kappa}{da} |_{\mathrm{LCP}} \langle \frac{\partial S}{\partial \kappa} \rangle_{sub.} 
  + a\frac{\partial j}{\partial a}|_{\mathrm{LCP}} \langle \frac{\partial S}{\partial j} \rangle_{sub.} \right).\nonumber\\ \label{eq:trace-anomaly}
\eeq
Here,  $a$ is the lattice spacing, and the beta-function for each parameter is evaluated at $\mu=T=0$ along the line of constant physics (LCP). Note that there is no renormalization for the quark number density as it is a conversed quantity.
We take all physical observables in the $j \rightarrow 0$ limit,  which implies that the third term in  the right side can be eliminated. 
$\langle \mathcal{O} \rangle_{sub.} (\mu) $ denotes the subtraction of the vacuum quantity. 
Thus, ideally, we should take $\langle \mathcal{O} \rangle_{sub.} (\mu) = \langle \mathcal{O} (\mu,T)  \rangle - \langle \mathcal{O} (\mu=0,T=0) \rangle $, but the exact zero-temperature simulations is practically difficult.
In this work, we take $\langle \mathcal{O} \rangle_{sub.} (\mu) = \langle \mathcal{O} (\mu, T=79\mathrm{MeV})  \rangle - \langle \mathcal{O} (\mu=0, T= 79 \mathrm{MeV}) \rangle $.

Utilizing the scale setting function (Eq.\ (23))  and a set of $(\beta,\kappa)$ with a fixed mass ratio of pseudoscalar and vector mesons $m_{PS}/m_V$ (Table~$1$)  in Ref.~\cite{Iida:2020emi}, 
 the coefficients can be nonperturbatively determined as 
 \beq
 a d\beta /da|_{\beta=0.80,\kappa=0.159}=-0.352, \quad a d\kappa/da |_{\beta=0.80,\kappa=0.159}=0.0282.\label{eq:beta-fn}
 \eeq


\section{Simulation results}
\begin{figure}[htbp]
 \begin{center}
    \begin{minipage}[b]{0.45\linewidth}
        \includegraphics[keepaspectratio, scale=0.6]{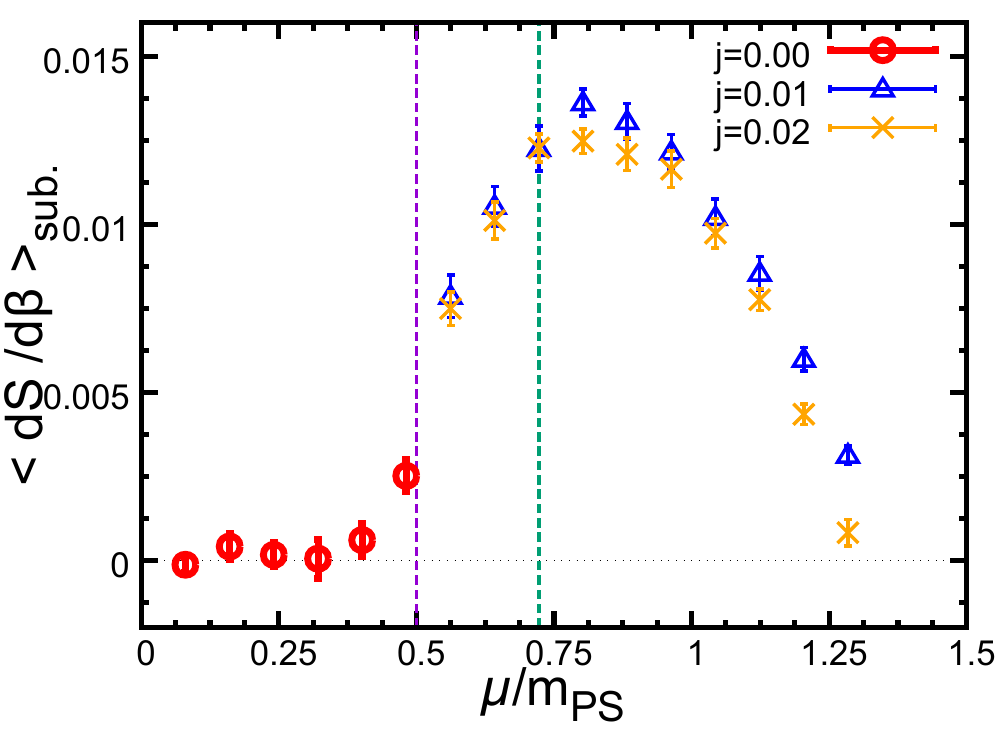}
    \end{minipage}   
    \begin{minipage}[b]{0.45\linewidth}
     \includegraphics[keepaspectratio, scale=0.6]{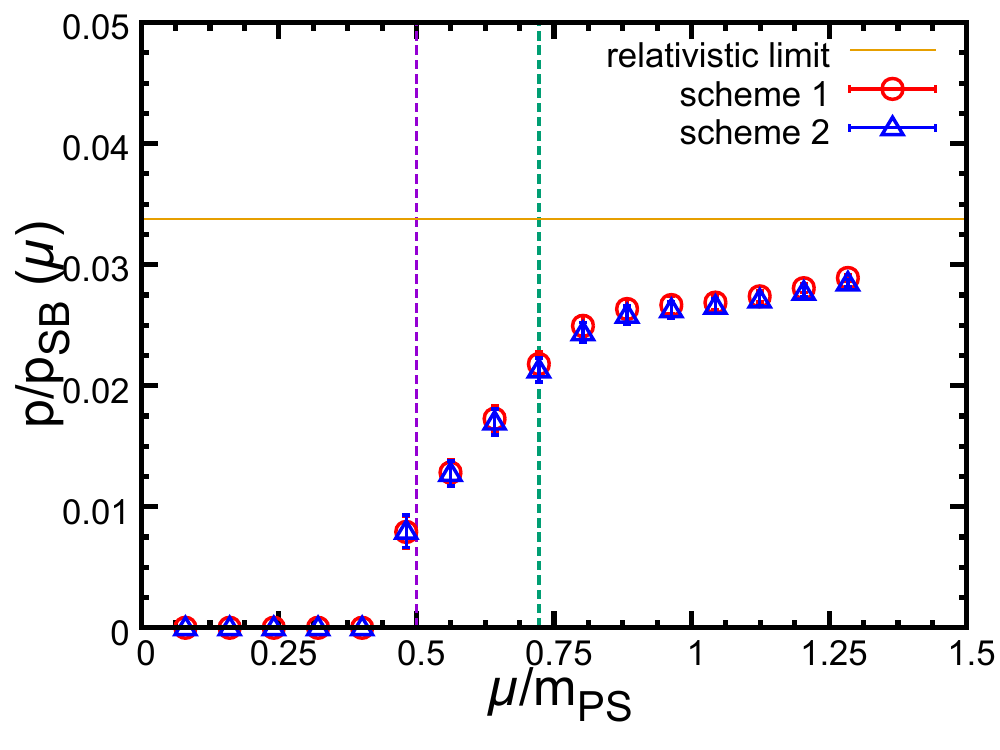}
     \end{minipage} 
    \caption{
(Left): Raw data of $\langle \partial S /\partial \beta \rangle _{sub.}$ for each $\mu$ and $j$. The purple dashed line denotes the critical value, $\mu_c$, which is the hadronic-superfluid phase transition point, while the green dashed line indicates that the BEC-BCS crossover occurs around this value of $\mu$.  (Right): Scheme dependence of the pressure.
}\label{fig:gauge-action}
\end{center}
  \end{figure}

The first term of the RHS in Eq.~\eqref{eq:trace-anomaly} is given by the measurement of the gauge action.
The raw data are plotted in the left panel of Fig.~\ref{fig:gauge-action}.
 We find that they are consistent with zero in the hadronic phase (except for near the phase transition point), while increasing in the BEC phase and then decreasing in the BCS phase.
Although we have determined the phase structure by the measurement of several physical observables in each phase as defined in Table~\ref{table:phase}, the results for $\langle \partial S /\partial \beta \rangle _{sub.}$ indicate that from the value of gauge action during the configuration generation, we can estimate where the hadronic-superfluid phase transition and the BEC-BCS crossover occur. 
Furthermore, we can see that the $j$-dependence is mild, so that we take the constant extrapolations of the $j=0.01$ and $j=0.02$ data in the superfluid phase.

The second term of the RHS in Eq.~\eqref{eq:trace-anomaly} is given by
\beq
\langle \frac{\partial S}{\partial \kappa} \rangle = \frac{1}{\kappa} \left( \mathrm{Tr}_{c, s,f}\mathbbm{1} - N_f \langle \bar{q}q \rangle \right).
\eeq
Thus, we measure the chiral condensate. To obtain the extrapolated value at $j=0$, we perform the reweighting of $j$ and take the linear extrapolation (see Fig.8 in Ref.~\cite{Iida:2019rah}).

The pressure can be expressed by the integral of the number density over $\mu$ in the thermodynamic limit.
On the lattice, two schemes with different discretization errors have been proposed in Ref.~\cite{Hands2006-mh}:
\beq
&\mathrm{Scheme~ I:} \frac{p}{p_{SB}}(\mu) &= \frac{\int_{\mu_o}^{\mu} d\mu' n_{q}^{latt.} (\mu') }{\int_{\mu_o}^{\mu} d\mu' n_{q}^{\mathrm{tree}} (\mu')},\\
&\mathrm{Scheme ~II:} \frac{p}{p_{SB}}(\mu) &= \frac{\int_{\mu_o}^{\mu} d\mu' \frac{n_{SB}^{cont.}}{n_{q}^{\mathrm{tree}}}   n^{latt.}_q(\mu')  }{\int_{\mu_o}^{\mu} d\mu' n_{SB}^{cont.} (\mu')}.\label{eq:p-scheme2}
\eeq
Here, $p_{SB}(\mu)$ denotes the pressure value at the Stefan-Boltzman (SB) limit, which is obtained by the numerical integration of the number density of quarks in the relativistic limit.
$\mu_o$ represents the onset scale, namely, the starting point at which $\langle n_q \rangle$ becomes nonzero as $\mu$ increases.
In the continuum theory, the pressure scales as $p_{SB}(\mu) = \int^\mu n_{SB}^{cont.}(\mu')d\mu' \approx N_fN_c \mu^4 /(12\pi^2)$ in the high $\mu$ regime, where $N_f$ ($N_c$) is the number of flavors (colors)

The simulation results are plotted in the right panel in Fig.~\ref{fig:gauge-action}.
First of all, we can see that the scheme dependence of $p$ is negligible. It indicates that the discretization effect of our simulation is small.
At $\mu_c=m_{PS}/2$ for the hadronic-superfluid phase transition (purple vertical line), $p$ takes  a nonzero value since  $\langle n_q \rangle$ becomes nonzero in the hadronic-matter phase.
Thus, $\langle n_q \rangle$ becomes nonzero before the hadronic-superfluid phase transition, then  $\mu_c$ is not the same as $\mu_o$. 
The low but finite temperature effects cause the discrepancy between them as discussed in~\cite{Iida:2019rah}. 
We can see that our data 
monotonically increase and approach the value in the relativistic limit.
The value of $p/p_{SB}$ is $\approx 0.84$ at the highest density in our simulation.

\begin{figure}[htbp]
 \begin{center}
    \includegraphics[keepaspectratio, scale=0.75]{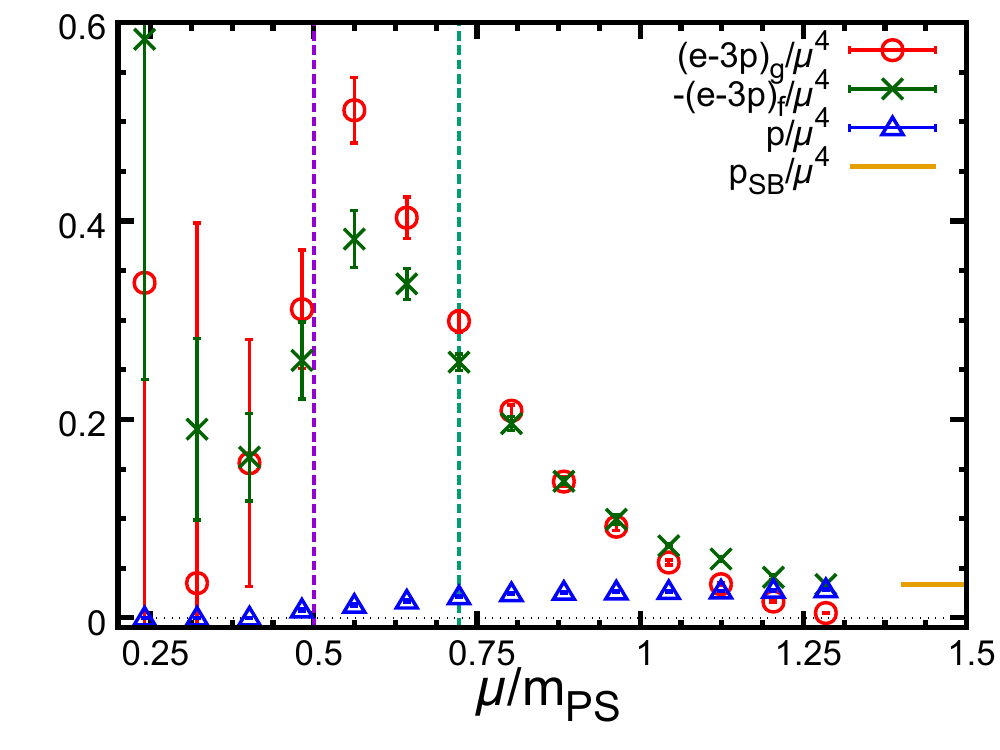}
    \caption{Trace anomaly and pressure  as a function of $\mu/m_{PS}$. The circle and cross symbols denote the gauge part and {\it minus} the fermion part of the trace anomaly, respectively. We also show $p/\mu^4$ at the relativistic limit, $p_{SB}/\mu^4=N_fN_c /(12\pi^2)$.
    }\label{fig:raw-data}
    \end{center}
  \end{figure}

The trace anomaly and pressure (Scheme II) are shown in Fig.~\ref{fig:raw-data}.
For the trace anomaly, we plot the gauge part (the first term in Eq.~\eqref{eq:trace-anomaly}) and minus the fermion part (the second term) separately.
Both parts are
normalized by $\mu^4$  
to see the dimensionless asymptotic  behavior.
The magnitude of each part has
a peak around the hadronic-superfluid phase transition. 
 It is very similar to the emergence of the peak of $(e -3p)/T^4$ around the hadronic-QGP phase transition at $\mu=0$.

 \begin{figure}[htbp]
 \begin{center}
    \begin{minipage}[b]{0.45\linewidth}
        \includegraphics[keepaspectratio, scale=0.6]{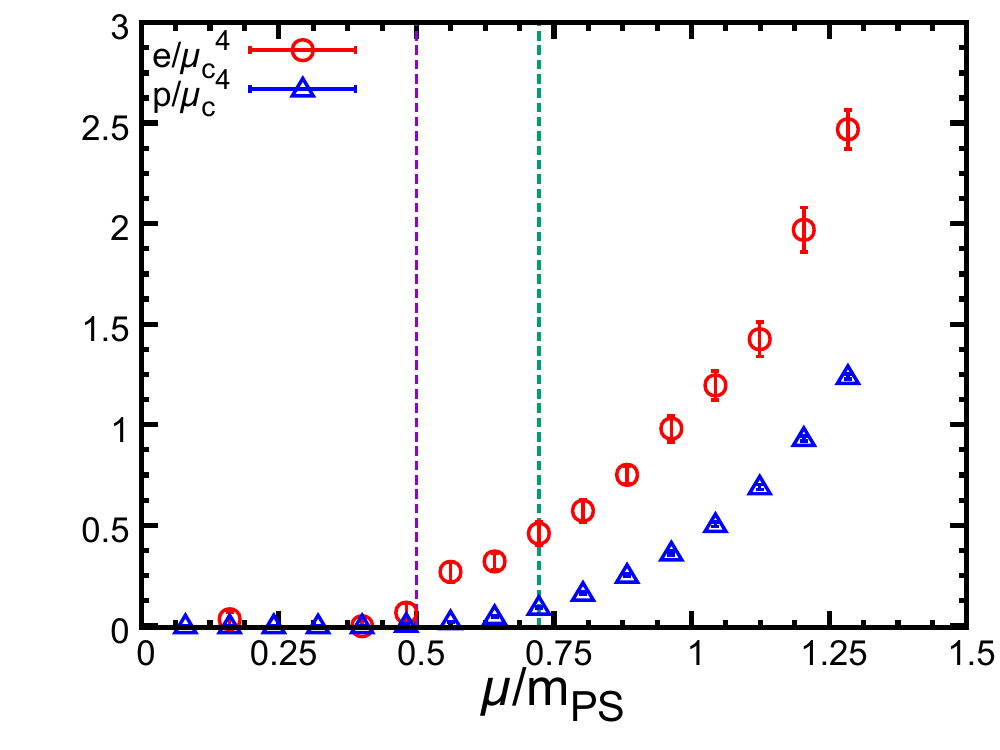}
    \end{minipage}   
    \begin{minipage}[b]{0.45\linewidth}
     \includegraphics[keepaspectratio, scale=0.6]{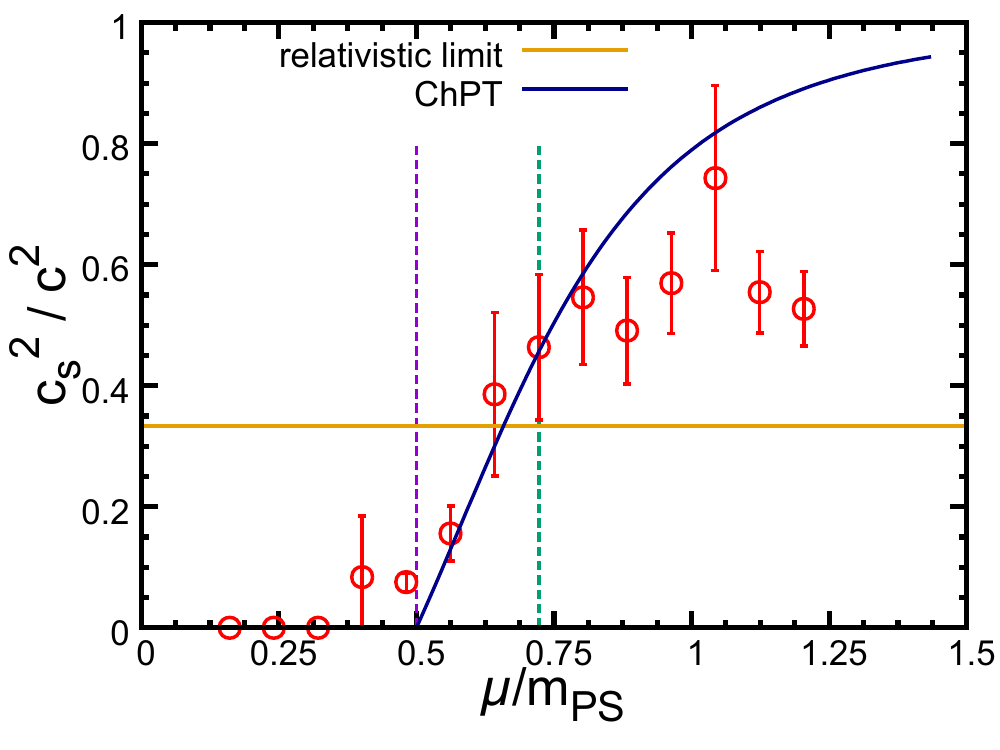}
     \end{minipage} 
    \caption{
(Left): The EoS as a function of $\mu/m_{PS}$.  (Right): Sound velocity  squared as a function of $\mu/m_{PS}$. The horizontal line (orange) denotes the value in the relativistic limit, $c_s^2/c^2 =1/3$. The blue curve shows the result of ChPT.
}\label{fig:EoS}
\end{center}
  \end{figure}
Combining the data of $e-3p$ and $p$ obtained above, we finally obtain the EoS and sound velocity in Fig.~\ref{fig:EoS}.
In the left panel, we normalize $e$ and $p$ by 
$\mu_c$ so as to be dimensionless. 
We can see that both $e$ and $p$ are consistent with zero in the hadronic phase.
Thus, these thermodynamic quantities are not changed even if $\mu$ increases before the hadronic-superfluid phase transition.
Note that $e\approx 0$ in hadronic phase indicates that the nonperturbative beta-functions of $\beta$ and $\kappa$ given by Eq.~\eqref{eq:beta-fn} work well enough to make the parts of trace anomaly, $(e -3p)_g$ and $(e -3p)_f$, cancel each other.

Now, let us focus on the sound velocity  depicted in the right panel in Fig.~\ref{fig:EoS}.
Here, we evaluate $c_s^2 (\mu)= \Delta p (\mu)/\Delta e (\mu) $, where $\Delta p (\mu)$ and $ \Delta e (\mu)$ are estimated by the symmetric finite difference, 
i.e., $\Delta p(\mu) =( p(\mu +\Delta \mu) - p(\mu -\Delta \mu))/2$.
First of all, our results are consistent with the prediction of ChPT~\cite{Son_2001, Hands2006-mh}, which is given by $c_s^2/c^2=(1-\mu_c^4/\mu^4)/(1+3\mu_c^4/\mu^4)$, in the BEC phase.
We also find that $c_s^2/c^2$ is larger than $1/3$, which is the value in the relativistic limit, 
at higher densities
than the regime where the BEC-BCS crossover occurs.
Eventually, our data  seem to peak around $\mu \approx m_{PS}$ and,  as density increases further, decrease so as to go away from the ChPT prediction.
{\it Such a peak of the sound velocity is a characteristic feature previously unknown from any lattice calculations for QCD-like theories.}
For example, in the finite temperature case, the sound velocity monotonically increases in $T>T_c$ and approaches 
 the relativistic limit as the temperature increases~\cite{Borsanyi:2013bia,HotQCD:2014kol}. 
 
Here, we give a comment on the holography bound. It is a conjecture that $c_s^2/c^2 \le 1/3$ is satisfied for a broad class of four-dimensional theories proposed by Ref.~\cite{Cherman:2009tw}.
The paper itself studies the finite temperature case in the context of holography. 
Our result from the first-principles calculation shows that the bound is broken in the case of finite density.
Furthermore, the counterexamples consisting of strongly coupled theories at finite density are also known in the context of holography~\cite{Hoyos:2016cob}.

\section{Summary and discussion}
It is strongly believed that at ultrahigh density, $c_s^2/c^2$ approaches the relativistic limit.
Then, there arises a question of
{\it how} it approaches $1/3$.  
According to the pQCD analysis (see Appendix~A in \cite{Kojo2021-on}), it scales as $c_s^2/c^2 \approx (1- 5\beta_0 \alpha_s^2/(48\pi^2))/3$, where $\beta_0 = (11N_c -2 N_f)/3$ denotes the $1$-loop coefficient of the beta-function.
Thus, $c_s^2/c^2$ approaches the asymptotic value from {\it below}. 
On the other hand, a result based on the resummed perturbation theory suggests that $c_s^2/c^2$ approaches 
the limit from {\it above}~\cite{Fujimoto2020-bh}. 
In the numerical simulations, the maximum value of $\mu$ is limited by $\mu \ll 1/a$ to avoid the strong lattice artefact. Otherwise, the hopping term of fermions would be partially suppressed by the factor $e^{-a\mu}$  in the Wilson-Dirac operator. For the extension to larger chemical potential, we need to perform smaller lattice spacing or lighter quark mass simulations. 
Furthermore, to obtain $c_s$ at $T=0$,
it is also required to see the EoS in the lower temperature regime by carrying out larger volume simulations.

According to Ref.~\cite{Kojo2021-mg}, a peak of $c_s^2$ appears due to the development of the quark Fermi sea just after the saturation  of low momentum quarks.
The density at which the peak appears in our results is apparently low, i.e., $\mu \approx m_{PS}$,
but seems sufficiently high that the quark Fermi sea would be fully developed.
It supports the predictions from several effective models  based on the presence of the quark Fermi sea~\cite{McLerran2019-qh, Kojo2021-mg, Kojo2021-wh}. 
Furthermore, it is reported that the peak of sound velocity emerges around BEC-BCS crossover also in condensed matter systems  with finite-range interactions~\cite{Tajima:2022zhu}.
To ask whether or not the emergence of the peak structure is a universal property of superfluids in a BEC-BCS crossover regime, it would be important to investigate the origin of this structure as 
another future work. 
If the peak of sound velocity would be a universal property even for real $3$-color QCD as discussed in Refs.~\cite{Kojo2021-mg, Kojo2021-wh}, then  it will change a  conventional picture 
that a first order transition from stiffened hadronic matter to soft quark matter is responsible for the presence of massive neutron stars.

\begin{acknowledgments}
We would like to thank T.~Hatsuda,  T.~Kojo, T.~Saito, D.~Suenaga, H.~Tajima and H.~Togashi for useful conversations.
We are grateful to S.~Hands and J.-I.~Skullerud for calling our attention to erroneous data in the earlier version of the manuscript.
The consistency with ChPT was kindly suggested by N.~Yamamoto.
E.~I. especially thanks T.~Kojo T.~Hatsuda and H.~Togashi for fruitful discussions about the origin of peak, the pQCD analysis and the correspondence between the lattice data and neutron-matter analysis.
Discussions in the working group ``Gravitational Wave and Equation of State" in iTHEMS, RIKEN was useful for completing this work.
The work of E.~I. is supported by JSPS KAKENHI with Grant Number 19K03875, JST PRESTO Grant Number JPMJPR2113 and JSPS Grant-in-Aid for Transformative Research Areas (A) JP21H05190,  and the work of K.~I. is supported by JSPS KAKENHI with Grant Numbers 18H05406 and 18H01211.
The numerical simulation is supported by the HPCI-JHPCN System Research Project (Project ID: jh220021).

\end{acknowledgments}

\bibliographystyle{utphys}
\bibliography{2color}

\end{document}